\documentclass[aps,prd,reprint,groupedaddress,amsmath,amssymb,superscriptaddress]{revtex4-1}

\usepackage{graphicx}
\usepackage{dcolumn}
\usepackage{bm}
\usepackage[utf8]{inputenc}
\usepackage[T1]{fontenc}

\begin{document}


\title{Do floating orbits in extreme mass ratio binary black holes exist?}


\author{Shasvath J. Kapadia}
\email[]{skapadia@uark.edu}
\affiliation{Department of Physics, University of Arkansas, Fayetteville, Arkansas 72701}
\author{Daniel Kennefick}
\email[]{danielk@uark.edu}
\affiliation{Department of Physics, University of Arkansas, Fayetteville, Arkansas 72701}
\affiliation{Arkansas Center for Space and Planetary Sciences, University of Arkansas, Fayetteville, Arkansas 72701}

\author{Kostas Glampedakis}
\email[]{kostas@um.es}
\affiliation{Departamento de Fisica, Universidad de Murcia, Murcia, E-30100, Spain}
\affiliation{Theoretical Astrophysics, University of T\"{u}bingen, T\"{u}bingen, D-72076, Germany}


\date{\today}

\begin{abstract}
This paper examines the possibility of floating or non-decaying orbits 
for extreme mass ratio binary black holes. In the adiabatic approximation, 
valid in the extreme mass ratio case, if the orbital flux lost due to 
gravitational radiation reaction is compensated for by the orbital flux 
gained from the spins of the black holes via superradiant scattering (or,
equivalently, tidal acceleration) the orbital decay would be stalled, 
causing the binary to ``float''. We show that this flux balance is not,
in practice, possible for extreme mass ratio binary black holes with
circular equatorial orbits; furthermore, adding eccentricity and inclination 
to the orbits will not significantly change this null result, thus ruling
out the possibility of floating orbits for extreme mass ratio binary 
black holes. We also argue that binaries consisting of material bodies 
dense and massive enough to generate gravitational waves detectable by any 
kind of gravitational wave detector are also unlikely to float. Using a 
multipolar analysis, we show that a non-Kerr spacetime  which could produce
a floating orbit (given the same amount of tidal acceleration as in the case
of a Kerr black hole) would need to be rapidly rotating prolate spheroid, which 
would be an exotic object indeed.
\end{abstract}


\maketitle

\section{Introduction}
The idea that a body, spiraling in towards a central spinning
black hole as a
result of gravitational wave damping of its orbit, might halt at a
certain radius while it absorbed energy from the central black
hole's spin, was first proposed by Charles Misner in 1972 \cite{misner72}, 
in
the context of the search for a plausible source for the claimed
gravitational wave detections of Joseph Weber \cite{weber70}. These constant 
frequency sources of gravitational waves were dubbed {\it floating
orbits} by William Press and Saul Teukolsky, who were the first to attempt
to find out if theory really predicted their existence \cite{pt72}. 
In the decades since several
researchers (at least) have attempted to find evidence for the 
existence of floating orbits in extreme mass ratio inspirals (EMRIs), in large part because extreme mass ratio binaries are expected to be good sources of low frequency gravitaional waves detectable to proposed space based interferometers like LISA which is built to be sensitive to waves of frequency range $1$ Hz to $10^{-4}$ Hz\cite{gkl05}. Floating orbits would emit gravitational waves at a constant frequency which would show up as peaks in frequency space standing out clearly from the background noise. However, none have published more
than a few lines in a longer report \cite{hughes01, gk02} relating a 
failure to find a floating orbit. More recently interest in the topic
has been reawakened by the discovery that an extreme-mass ratio binary
with a floating orbit is possible where a massive scalar field mediates
the energy exchange \cite{cardetal}.
The current paper seeks to show
as conclusively as possible that the original floating orbit envisaged
by Press and Teuksolsky can now be ruled out for extreme-mass ratio
binaries.

The mechanism identified by Press and Teukolsky as the one by which 
the orbiting body in an extreme mass ratio inspiral can steal the central black hole's
rotational kinetic energy is that of {\it superradiant scattering} \cite{pt72}. In superradiant scattering, the gravitational waves reflected from the central spinning body are more energetic than those impinging upon it, the extra energy coming at the expense of the spin of the central body. These waves thus carry some of the stolen spin energy to the orbiting body, which, upon absorbing it, gains orbital energy.  

It has been proposed \cite{membrane} (though, to our knowledge not formally proven in general) that the superradiant
scattering picture of this process is functionally equivalent to
the process of tidal friction between orbiting black holes. First developed by Hartle for a particle orbiting a slowly rotating black hole at a large orbital radius \cite{hartle73} (a situation for which tidal friction can be formally shown to be equivalent to superradiant scattering), this mechanism is familiar from Newtonian orbital mechanics
between material bodies, for instance the process by which the Moon
extracts rotational kinetic energy from the Earth in order to move
away from us as it gains orbital kinetic energy. Analogously one can
visualise the orbiting body distorting the event horizon of the 
central black hole, effectively raising tidal bulges, and then 
interacting with those bulges in order to torque the black hole,
braking its rotation, while accelerating the orbiting body itself.

Most methods of calculating the energy interchange between the
two bodies in an extreme mass ratio inspiral have made use of the Teukolsky formalism \cite{teukolskyI73, teukolskyII73, teukolskyIII74} to
calculate the energy lost from the large central black hole. Recently
Poisson has presented calculations based on the tidal-friction picture
which give the energy extracted from the smaller body if it too is
a black hole \cite{poisson04, poisson09}. Clearly the best possible chance of finding a floating
orbit is if the two objects in the binary are both spinning and
are orbiting each other at a small radius, while
each extracts rotational energy from the other and converts it into
orbital kinetic energy. Of course in practice we would expect the
smaller body in an EMRI which had slowly evolved to a small radius
to have become tidally locked to the larger body, all spin energy
having been extracted (though if its orbit was eccentric, as it
would probably be, then the eccentricity of the orbit would provide
further extractable energy). Additionally in an EMR inspiral there
is just a huge difference in the energy stored in the small body
compared to the large body. Still, it is interesting to look at the
best possible scenario before closing the door on floating orbits
in extreme mass ratio (EMR) black hole binaries once and for all.

It should also be kept in mind that though floating orbits may not
exist, the energy interchange between the two bodies can still play
a significant role in the evolution of the orbit, and since this in
turn is critical to data analysis in gravitational wave detection,
it is certainly of interest to see what needs to be calculated and
what does not \cite{price}.

The general program adopted by the paper goes as follows:
EMR binaries in circular equatorial orbits are considered, with 
radius $R$ and hole masses $m$ and $M$.  
The orbital flux gained from the larger black hole is calculated using 
the Teukolsky (superradiant scattering) formalism using a code
described in \cite{gk02}, and compared with 
the orbital flux radiated away. The flux gained turns out to be 
less than $10\%$ of the flux lost.  

The orbital flux extracted from the smaller black hole is calculated using Poisson's 
results (tidal friction formalism) \cite{poisson04, poisson09}, and found to be about several orders of magnitude smaller than the orbital flux radiated away.
There is a denominator in the term describing
the energy loss by the smaller black hole in this formalism which
would see the energy loss diverge. This might conceivably be of physical
interest except that the divergence takes place only at a radius inside
the innermost stable circular orbit (ISCO) of the orbiting system. Thus, beyond that point, not only is no stable orbit possible, but the approximation scheme loses its validity \cite{poisson09}.

And finally, the fluxes gained from both black holes are added, and compared with the flux lost. The contribution to the orbital flux from the smaller hole being insignificant in comparison to the contribution from the larger hole, the total orbital flux gained remains less than $10\%$ of the orbital flux radiated away. We conclude 
that floating orbits for EMR binary black holes moving in circular 
equatorial orbits are not possible. 

Eccentric and non-equatorial orbits are then considered, and 
plausible arguments are given as to why such orbits cannot float. The case where the black holes are replaced
by material bodies is also explored. For EMR binaries this is unlikely
to produce floating orbits in any system which is producing detectable
gravitational waves.

To further support our claim that EMR binary black holes probably don't float, we employ the flux balance approach outlined above and equate an expression for the flux radiated away to infinity found in Ryan's multipolar formalism \cite{Ryan97} with an approximate post-Newtonian expression for the horizon flux. We thus infer the degree to which the spacetime needs to be ``non-Kerr'' in order to allow floating orbits to exist and conclude that either the spinning body must be prolate (which is contrary to the general expectation), or that the orbital flux extracted from the spinning black hole must be greater than the flux extracted via superradiance/tidal friction in the Kerr case.

Geometrized units $(G = c = 1)$ are used throughout the paper.

\section{Gain in orbital energy from the spin of the larger black hole} \label{superradiant}

\begin{figure}[htb]
\begin{center}
\includegraphics[width=0.5\textwidth]{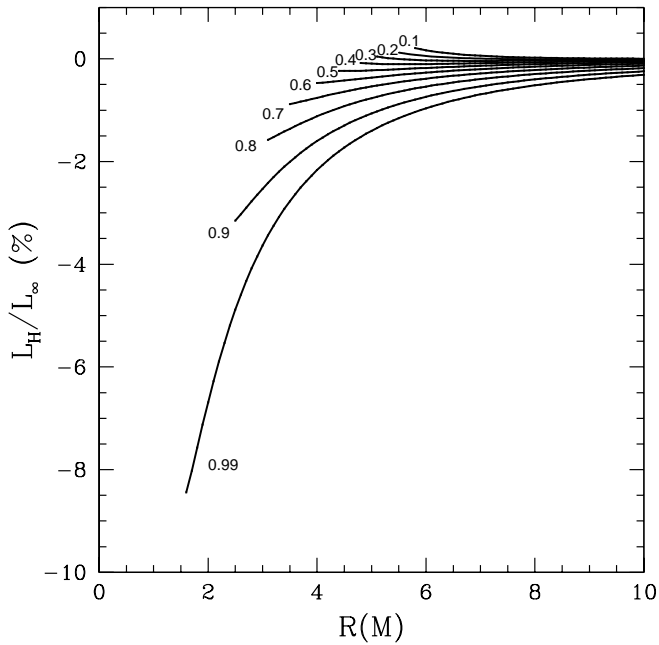}
\end{center}
\caption{Plot of the ratio $L_{H}/L_{\infty}$, comparing the flux of
gravitational waves arriving at the horizon of the larger black hole to
the flux reaching infinity,  
versus the orbital radius $R$, for different Kerr parameter values $a$ of 
the larger black hole. $|L_{H}|$ is at best about 8\% (for $a = 0.99$) of $L_{\infty}$, which occurs when $a = 0.99$ and $R = R_{\mathrm {ISCO}}$. A floating orbit requires a ratio of 100\%.}
\label{fig:FluxRatio}
\end{figure}

For EMR binaries, the gravitational wave flux emitted to infinity $L_{\infty}$ 
by the binary and the flux emitted by the smaller non-spinning black hole towards the 
horizon of the larger hole $L_{H}$ may be calculated using the Teukolsky 
formalism \cite{teukolskyI73, teukolskyII73, teukolskyIII74} (see Appendix~\ref{Tformalism}). If $L_{H} < 0$, superradiant scattering is occuring 
and the binary is gaining orbital energy at the expense of the spin energy 
of the larger black hole. We define the superradiant flux $L_{S}$ as equal to $L_{H}$ when $L_{H} < 0$.
Newtonian or post-Newtonian order approximations tend to suggest that there
is a wide gulf between the magnitudes of $L_S$ and $L_\infty$ but for orbits close to the
ISCO more exact calculations using the code described in \cite{gk02} show that
the two quantities can come close without ever quite 
producing a floating orbit.
Figure~\ref{fig:FluxRatio} plots the ratio of the fluxes to the horizon and infinity as a function of orbital radius. Note that though the formalism requires that the mass ratio be much less than unity, the results in this section are independent of its precise value, since we deal with the ratio of two fluxes which have the same dependence on the mass ratio.

\begin{figure}[htb]
\begin{center}
\includegraphics[width=0.5\textwidth]{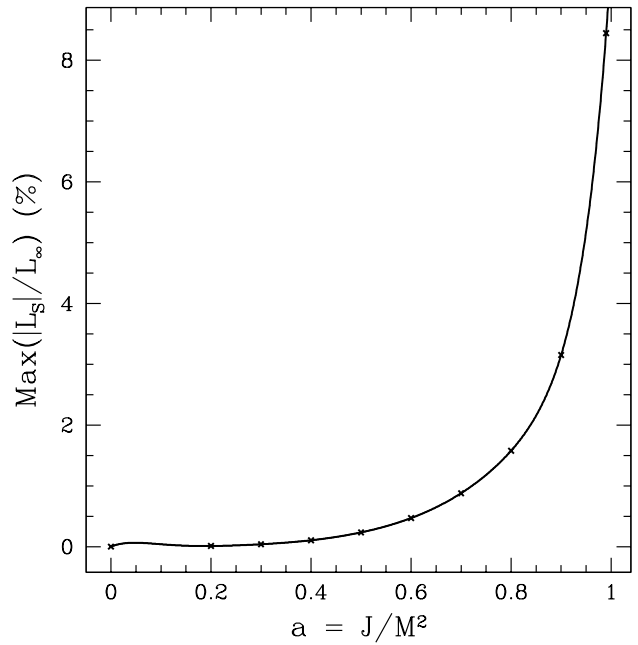}
\end{center}
\caption{Plot of the maximum superradiant flux ratio for each value of the spin parameter $a$ corresponding to Figure ~\ref{fig:FluxRatio}, as a function of $a$. The extrapolated estimate for the magnitude of the maximum possible superradiant flux ratio is close to 10\%.}
\label{fig:MaxFlux}
\end{figure}

Figure~\ref{fig:FluxRatio} suggests that the flux ratio increases with increasing Kerr parameter value. Figure \ref{fig:MaxFlux} shows the maximum superradiant flux ratio calculated for different values of the Kerr parameter $a$, and plotted as a function of $a$. The curve is then extrapolated to $a = 1$ to determine the maximum possible flux ratio.

It is clear from Figure \ref{fig:MaxFlux} that the maximum possible flux ratio does not exceed $10\%$, 
comfortably less than the required $100\%$ for a floating orbit. Thus, the orbital flux 
gained from the spin of the larger black hole, calculated using superradiant scattering, 
is not sufficient to generate a circular equatorial floating orbit. 

\section{Gain in orbital energy from the spin of the smaller black hole} \label{tidal}
The code based on the Teukolsky formalism, used to generate data for superradiant scattering found in the previous section, is ineffective in determining the orbital flux gained by an EMR binary black hole from the spin of the smaller black hole. There is however a way around this; since, as mentioned in the Introduction, tidal friction and superradiance are just different ways of describing the same unique process that converts spin energy to orbital energy, we may use the former to determine the orbital flux gained from the spin of the smaller black hole. 

\begin{figure}
\begin{center}
\includegraphics[width=0.5\textwidth]{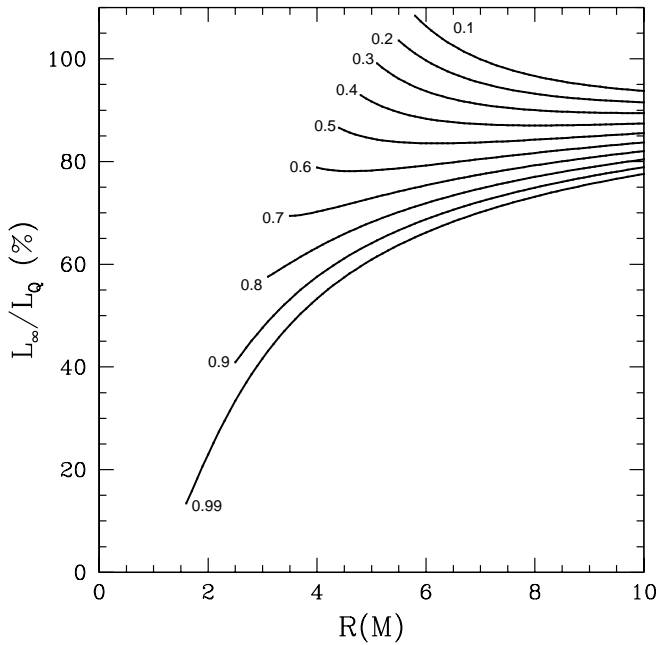}
\end{center}
\caption{ Plot of the ratio $L_{\infty}/L_{Q}$ versus $R$, for different spin parameter values $a$. The figure suggests that $L_{\infty}/L_{Q}$ tends to $100\%$ asymptotically with increasing orbital radius. However, for orbits close to the ISCO, the quadrupole formula becomes increasingly inaccurate, with $L_{\infty}/L_{Q}$ dropping below $20\%$ near the ISCO radius for $a = 0.99$. }
\label{fig:quadLinfRatio}
\end{figure}

The works of Poisson et al \cite{poisson04, poisson09}  give us analytical expressions for the spin down of the black holes via tidal friction in various limiting situations such as extreme mass ratio ($m \ll M$) and large orbital radii ($R \gg M$). These may then be converted to expressions for the orbital flux gained ($L_{t}$), and compared to the orbital flux emitted to infinity. It would be useful to have an analytical expression for the orbital flux radiated away, allowing for an algebraic approximation for the floating orbit radius. The quadrupole formula \cite{petersandmatthews}: 
\begin{equation}\label{quad}
L_{Q} = \frac{32}{5}\frac{m^2}{M^2}v^{10}
\end{equation}
where $v = \sqrt{\frac{M}{R}}$ the orbital velocity, is a Newtonian order approximation for $L_{\infty}$ (sufficient for the order of magnitude estimates in this section), which becomes increasingly accurate with increasing orbital radius, as shown in Figure \ref{fig:quadLinfRatio}. 

Equating $L_{Q}$ with $L_{t}$ and solving for $R$ gives an estimate of the orbital radius $R_{f}$. Using the expressions available from the works of Poisson et al \cite{poisson04, poisson09}, it is clear that the orbital flux gained from the spin of the smaller black hole is maximized when both black holes are spinning, with no restriction on the Kerr parameter of either black hole.

From \cite{poisson09} (see Appendix~\ref{Tfriction}), 
\begin{equation}\label{energy4}
L_{t} = \frac{8}{5}\frac{m^{5}}{M^{5}}b(1+3b^{2})v^{15}\Gamma_{K}
\end{equation}

where $b$ is the spin parameter of the smaller black hole, and $\Gamma_{K}$ is
a relativistic factor \cite{poisson09} (see Appendix~\ref{Tfriction}). It turns out that the relativistic factor becomes independent of $a$ in the limiting case when the velocity parameter tends towards its ISCO value \cite{poisson09}:

\begin{equation}\label{gammab}
\Gamma_{K}(a,b,v_{\mathrm{ISCO}}(a)) = \frac{5}{9}\frac{4+9b^2}{1+3b^2} \equiv \Gamma_{K}(b)
\end{equation}

The maximum value that the relativistic factor can take is $\Gamma_{K}(b = 1) = 65/36$. Using this value, 
the maximum value that $L_{t}$ can take is found to be $L_{t}^{max} = (104/9)\times(m/M)^5$, 
when $a = 1$ and $v = v_{\mathrm{ISCO}}(a = 1)$. At the ISCO of an extremally spinning black hole $(a = 1)$, the 
quadrupole formula assumes the value $L_{Q} = 32/5 \times (m/M)^2$ (cf equation \eqref{quad}). Because $L_{t}$ and $L_{Q}$ have different dependencies on the binary's mass ratio, we select the classic EMRI ratio of $10^{-6}$ representing a stellar mass black hole ($\sim 10M_{\odot}$) orbiting a supermassive black hole ($\sim 10^7M_{\odot}$).

$L_{t}^{max}$ scales as $(m/M)^5 \sim 10^{-30}$, eighteen orders of magnitude less than the quadrupole 
formula $L_{Q}$, which scales as $(m/M)^2 \sim 10^{-12}$. Thus we see, not surprisingly, that the trouble
in the EMRI case is that the smaller hole simply does not possess enough spin energy to stall the orbital decay. The formalism used in this paper can only deal with the EMRI case but this does give reason to believe
that in an equal-mass binary with both black holes spinning that the orbital energy gain would have a
strong effect on future orbital evolution, even if a floating orbit did 
not occur, as argued in
\cite{price}.

\begin{figure}
\begin{center}
\includegraphics[width=0.5\textwidth]{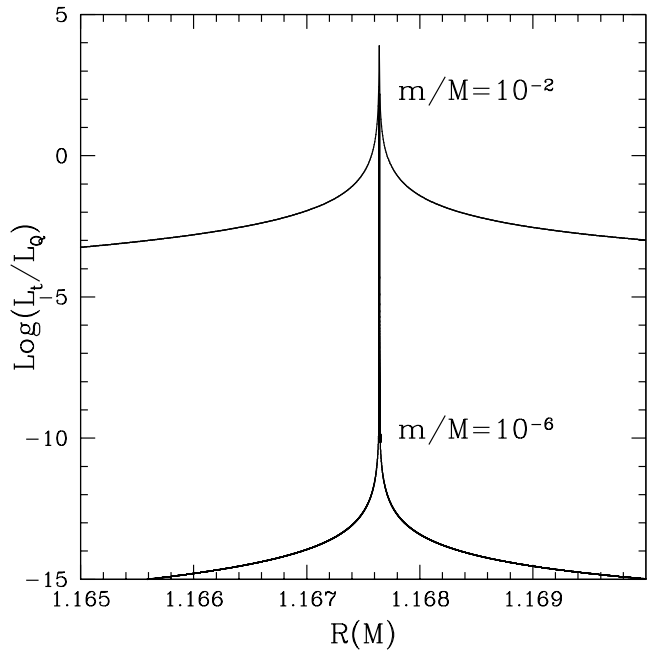}
\end{center}
\caption{Plot of the logarithm (base 10) of the ratio of the orbital flux gained $L_{t}$ to the flux radiated away, approximated here with the quadrupole formula $L_{Q}$, for mass ratio $m/M = 10^{-6}$ and $m/M = 10^{-2}$. At $R = 1.16765 M$, $L_{t}$ diverges and thus becomes greater than $L_{Q}$. But, for both mass ratios, this occurs inside the ISCO at $R_{\mathrm{ISCO}} = 1.45M$, where the approximation is invalid and where no stable orbit can physically exist. Given the small effect that changing the mass ratio by four orders of magnitude has on the radius where both fluxes $L_{t}$ and $L_{Q}$ equal each other, we don't expect the above result to change significantly for a mass ratio of even $m/M = 10^{-1}$. Higher mass ratios will tend to strain the EMR approximation.} 
\label{fig:blowup99}
\end{figure}

Given the enormous difference in scaling between \eqref{energy4} and \eqref{quad}, the only hope for a floating
orbit would come if the  
denominator $(1-3v^{2}+2av^{3})^{2}$ of the ``relativistic factor'' $\Gamma_{K}$ in \eqref{energy4} (see equation \eqref{gammaVba}) could drop to a value close to zero: 
(An example of this is shown in Figure:~\ref{fig:blowup99} for $a = 0.99$). One would still have to determine
if such behaviour was physical, but in any case it turns out that this equation diverges only at a
radius which is inside the ISCO, and thus no stable orbit is possible with this property. 

Furthermore, we illustrate, in Figure:~\ref{fig:blowup99}, the case of a mass ratio of order $1/100$, which is probably close to the limit of applicability of our approximation schemes. We see that, at this mass ratio, we are still not close to a floating orbit. If the mass ratio were increased further, to order $1/10$, the flux ratio $L_{t}/L_{Q}$ would still be of order $10^{-3}$, not nearly enough to sustain a floating orbit.

\begin{figure}
\begin{center}
\includegraphics[width=0.5\textwidth]{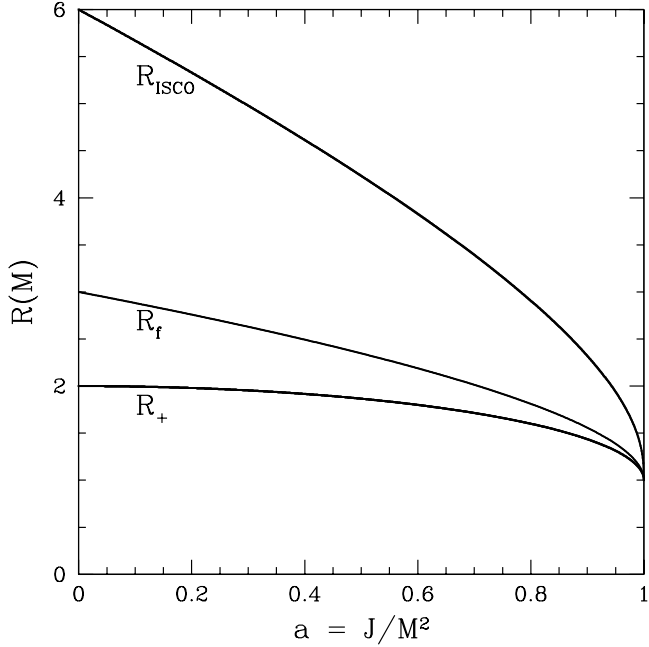}
\end{center}
\caption{Plot of the event horizon radius $R_{+}$, ISCO radius $R_{\mathrm{ISCO}}$, and the radius $R_{f}$ at which the orbital flux extracted from the smaller black hole and the flux radiated away equal each other, as a function of $a$. Clearly, $R_{+} < R_{f} < R_{\mathrm{ISCO}}$, except for the physically irrelevant spin parameter value $a = 1$ for which $R_{+} = R_{f} = R_{\mathrm{ISCO}} = M$.} 
\label{fig:poisson09}
\end{figure}

Figure \ref{fig:poisson09} clearly shows that for all values of the Kerr parameter $a$ of the larger black hole, 
the possible floating orbit radius (created by the putative divergence in
the tidal friction energy-extraction formula) 
does not exceed the radius of the ISCO. However, for the case of extremal spin 
($a = 1$), the floating orbit radius equals both the ISCO radius and the event horizon 
radius ($R_{+} = R_{f} = R_{\mathrm{ISCO}} = M$). Since the value $a=1$ is not achievable physically, and buffering
probably sets an upper limit on black hole spin of $a<0.998$ \cite{ThorneBuffer} this is not a caveat
with any physical relevance. 


Using the numerical value of the flux radiated to infinity rather than its algebraic approximation will not change this result, since $L_{\infty}/L_{Q} \sim 1$ and $L_{t}/L_{Q} \sim (m/M)^3 = 10^{-18}$ (for $m/M = 10^{-6}$), which means $L_{t}/L_{\infty} \sim (m/M)^3 = 10^{-18}$, far less than the required $100\%$ for a floating orbit.

It seems clear that, a circular equatorial floating orbit generated using the orbital energy extracted from the spin of the smaller black hole, calculated using the tidal friction formalism, is not possible.

\section{Adding contributions from both black holes to the gain in orbital flux}

Neither black hole can, by itself, stall the orbital decay. Evidently, adding both these gains in orbital energy maximizes the possibility of a floating orbit. The total orbital flux gained is defined as:

\begin{equation}
L_{T} = |L_{S}| + L_{t}
\end{equation}

A comparison of the fluxes extracted from the larger and smaller black holes shows that $|L_{S}|$ is several orders of magnitude greater than $L_{t}$. $|L_{S}|/L_{\infty} \sim 10^{-2}$ and $L_{t}/L_{\infty} \sim 10^{-18}$, and therefore $L_{t}/|L_{S}| \sim 10^{-16}$. Clearly, $L_{T} \simeq |L_{S}|$. 


It was previously shown that $|L_{S}|/L_{\infty} < 10\%$. Therefore, $L_{T}/L_{\infty}$ will certainly not exceed this percentage either.

Changing the mass ratio (as done in the previous section) to something as high as $10^{-1}$, the approximation schemes used so far would still be some guide (though not completely reliable). The flux ratio $|L_{S}|/L_{\infty}$ would remain unchanged, but the flux ratio $L_{t}/L_{\infty}$ would still be small, i.e of order $10^{-3}$. 
In fact it seems that the comparable mass ratio regime is, in practice, even
less favorable to floating orbits than the extreme mass ratio regime. Published
results suggest that there is relatively little tidal torquing associated
with equal mass binaries, since the inspiral is associated with very 
minor changes in the mass of the black 
hole. \cite{Alvi, Lovelace}


The analysis in this and the previous sections strongly suggests that EMR binary black holes in circular equatorial floating orbits cannot float.

\section{Other Scenarios}

It is natural to inquire whether a floating orbit might still be 
possible outside of the restricted
scenario studied here. The extreme mass ratio of the binary is an assumption which is integral to our
approach. Probably full numerical relativity will be required to answer the question of whether floating
orbits are possible for comparable mass binaries, though it seems unlikely. But several other scenarios
can be discussed with interest.

\subsection{Eccentric Orbits}

Might eccentric orbits be more likely to be of the floating type? If we consider the case of Newtonian
tidal friction in systems involving planets and moons
it is true that orbital eccentricity provides another pool of
energy (besides rotational energy of a planet's spin) which can be transformed into orbital energy.
Tidal friction ceases to operate on a moon when it becomes tidally locked to its planet, (when the moon always presents the same face to its planet) but this cannot
happen if its orbit is eccentric (instead the orbiting body, as in the case of
Mercury, enters a resonance state in which its rotation is coordinated with
its orbital motion but not actually locked to it), so tidal friction can also operate to drive an orbit more circular,
extracting a little energy for increasing the orbital radius at the same time. 

\begin{figure}
\begin{center}
\includegraphics[width=0.5\textwidth]{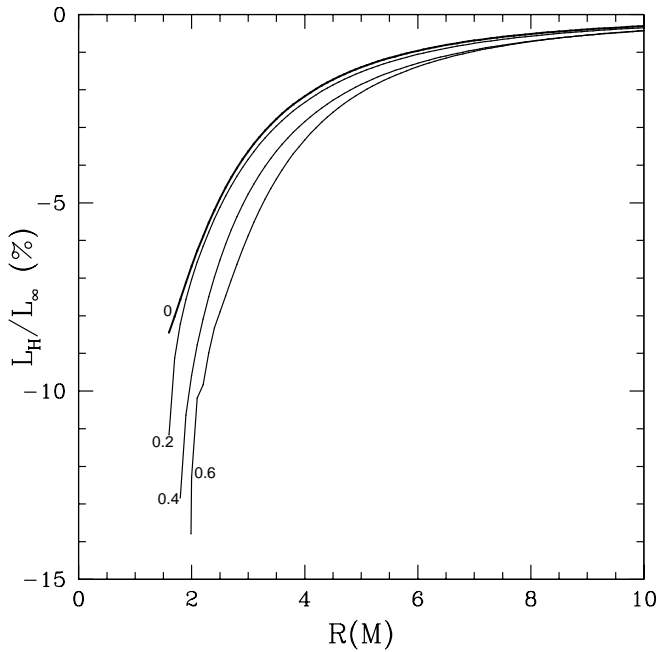}
\end{center}
\caption{Plot of $L_{H}/L_{\infty}$ for various eccentricities $e$ and $a = 0.99$, as a function of the semi latus rectum $R$. This ratio, and thus the superradiant flux ratio $|L_{H}|/L_{\infty}$, is within 20\% for all eccentricities used.} 
\label{fig:eFluxRatio}
\end{figure}

The code discussed in \cite{gk02} can deal with orbits of moderate eccentricity and we show in Figure \ref{fig:eFluxRatio}
that eccentricties up to $e=0.6$ only slightly increase the rate of orbital energy gain by the orbiting
body. 

What if we could deal with arbitrary eccentricities? 
First it is important to note that an orbit with very high eccentricity
close to the ISCO is quite unlikely, since tidal damping, which is the
usual method by which orbits close to the ISCO are generated, tends to
damp out eccentricity. Studies show that, close to the ISCO EMRIs will 
typically have eccentricities $e<0.6$ \cite{gkl05}. 

Nevertheless one cannot rule out the
possibility of a chance capture of the smaller black hole in just such
an unlikely ISCO skimming orbit. In such a case it would be best placed
to transfer the energy associated with its orbital eccentricity to the
task of expanding the orbit. Although one cannot accurately calculate
the rate of energy transfer for such a high eccentricity orbit,
one thing we can do is to look at the
total energy reservoir available for such a transfer. Essentially it lies in the difference between the
energy of the eccentric orbit and that for an "equivalent" circular orbit, defined as the orbit with the same perigee but zero eccentricity. Recall
that tidal friction in the solar system occurs not only in cases, such
as the Earth-Moon system, where a body retains angular momentum, but also
in cases of tidally locked bodies, such as Io, which retain some eccentricity
in their orbit.
If one plots
the effective potential, for a given orbital angular momentum, of a particle orbiting
a Kerr black hole, one can easily read off the difference in energy between orbits of differing
eccentricities. It can be shown (Table~\ref{ecctable}) that the energy differential between a 
near-parabolic and a circular orbit
is not that significant compared to the total orbital energy and that the difference in energy between
a parabolic orbit and one of $e=0.6$ is much less again.
\begin{table}[b]
\caption{\label{ecctable}The table shows values of eccentricity ($e$) and the corresponding specific energy ($E$) of the smaller black hole orbiting the larger black with $a = 0.99$. There is only a $35\%$ increase in specific energy when going from a circular ($e = 0$) to a highly eccentric ($e = 0.99$) orbit.
}
\begin{ruledtabular}
\begin{tabular}{cc}
\textrm{e (eccentricity)}&
\textrm{E (specific energy)}\\ 
\colrule
0 & 0.73597\\
0.1 & 0.73597\\
0.3 & 0.79123\\
0.6 & 0.87330\\
0.99 & 0.99657
\end{tabular}
\end{ruledtabular}
\end{table}

Since if the total amount of extra energy available in a very eccentric orbit
is small, it seems hardly likely that the rate of energy
transfer will be significantly augmented. It seems unlikely that anyone will find an eccentric orbit
which can float.

\subsection{Inclined Orbits}

An inclined orbit might provide another way in which some energy could be extracted for a floating
orbit. Although inclination can be altered by radiation damping, Hughes has shown that increasing the inclination angle only decreases the orbital flux extracted from the spin of the central black hole \cite{hughes01}, seemingly closing the possibility of circular inclined floating orbits . 
It would be surprising therefore, if inclined eccentric orbits floated when inclination by itself does not help and eccentricity seems to help
only marginally.

\subsection{Material bodies}

Based on our numerical results, the maximum superradiant flux ratio $\mathrm{Max}(|L_{S}|/L_{\infty})$ 
increases with increasing  Kerr parameter value. For black holes, 
there are limitations on $a$ , both in principle ($0 < a <1$) and in practice
due to buffering ($a<0.998$) \cite{ThorneBuffer}. 
However, the limitations are not the same for all astrophysical objects, and it is certainly of academic interest to ask whether floating orbits are possible around other astrophysical bodies.

\begin{figure}
\centering
\includegraphics[width=0.45\textwidth]{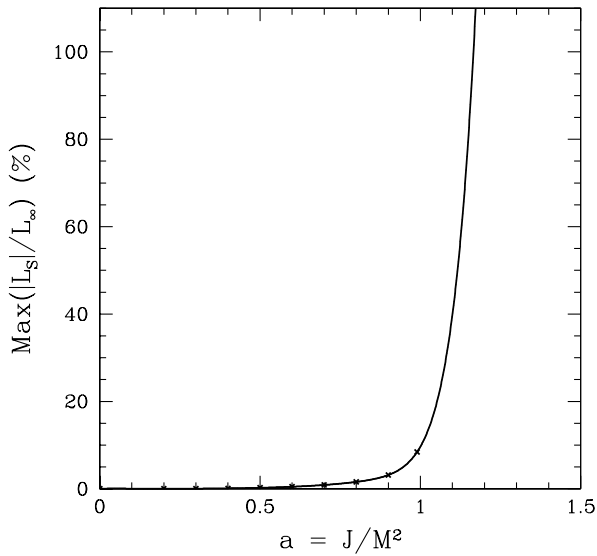}
\caption{Same as Figure \ref{fig:MaxFlux}, extrapolated to a maximum superradiant flux ratio of 100\% at $a \simeq 1.2$}
\label{fig:FluxRatio100}
\end{figure}

It is interesting to speculate on which value of $a$ would, extrapolating
from our results, seem likely to permit a floating orbit (Figure \ref{fig:FluxRatio100}). 

Obviously our results, which depend on the assumption that we are dealing
with the Kerr spacetime, are formally invalid for values of $a>1$. 
Nevertheless it is highly plausible that increasing the angular momentum
of the larger body would make a floating orbit more likely and we even
have some evidence of this in the form of real life systems which actually
float. One example of such a system is the one in which we live, the
Earth-Moon system (note that the Kerr parameter of the Earth is $a=732$). 
The Moon extracts more energy from the Earth's spin
than it loses due to the emission of gravitational waves, and consequently
retreats from us steadily, rather than inspiralling towards us.
In fact, we are completely unable to even
measure the effects of gravitational wave emission (if it
exists) from this system. But the Earth-Moon system is radically
different from the kinds of systems which are of interest to
gravitational wave astronomers, because its orbital velocities are very small 
while the spin parameter of the Earth is very high. A system which can
maintain very high orbital velocities, and thus emit significant gravitational
radiation, must contain compact massive objects and there is evidence that
such bodies cannot sustain large spin parameters.

It is worth mentioning here that an alternative channel for tidal dissipation, 
involving scalar fields, has been discussed recently
\cite{Cardosoetal, cardetal}. They propose that a floating orbit may
be possible presuming the existence of massive scalar fields which, however,
are also known to lead to instabilities \cite{detweiler80}.

One question of interest is whether both viscosity-based
tidal dissipation
(of the Earth-Moon type) and superradiance-based tidal damping
can exist in the same system. Obviously if both objects are
black holes then only the latter type is possible, since there
is no matter with non-zero viscosity involved. A hyper-compact
object might, at least at first glance, appear to be capable
of both. The usual model for such a system is a Kerr spacetime
with a mirror capable of reflecting gravitational waves placed
in front of the horizon (on the grounds that the waves will otherwise pass
right through a material body and re-emerge on the other side).
Note, however, that 
Richartz et al \cite{Richartzetal} show that replacing the event
horizon with a mirror in 
this
way generally destroys superradiance in a system. 


Replacing either of the black holes in our EMRI system with a material
body of some sort will only help us achieve a floating orbit if we do
so with the larger of the two bodies (the extrapolation above was naturally
based upon this assumption). For such a system to emit detectable
gravitational waves the larger body should be a
supermassive black hole (or at least intermediate mass,
depending on its distance from our detectors and their
sensitivity) and it is not at all clear that material bodies exist with masses
above several hundred solar masses. Of course floating
orbits could exist in comparable mass binaries, which
would be of interest to LIGO, but this is beyond the
scope of our paper to analyze. Beyond that, while
tidal friction is important (presumably) in any material body, for
gravitational waves to be important in the evolution of the system,
the material body substituted for the black
hole must be a hyper compact body, at least as compact
as a neutron star. Most likely we are looking for an object
which is compact enough to have an ergoregion, but not
compact enough to have an event horizon. Whether such objects exist 
remains to be seen, but many candidates have been proposed (including 
boson stars,
gravastars and others), though many may be subject to
instabilities \cite{Cardosoetal08}, although in some cases the
ergoregion instability can be very long lived \cite{ChirRez}.



The most
compact object of which we have knowledge is the neutron star. But
theory suggests that the largest value of $a$ possible for a neutron
star is $0.7$ \cite{lo11} and the neutron star has no ergoregion. Thus it
seems highly unlikely that a neutron star binary can have a floating
orbit. In any case, neutron stars cannot have more than a few solar
masses, which makes them irrelevant to our EMRI case. A quark star,
on the other hand, if such a thing exists, can theoretically have 
$a$ of up to $1.2$ \cite{lo11} and would have an ergoregion, but, in all likelihood, they cannot be of the supermassive type, according to current theoretical models. Boson stars may have a spin parameter of up to 1.4 \cite{Ryan97} and are compact enough to have an ergoregion. This makes them ideal candidates for a
floating orbit. Gravastars are also candidates, partly because so little is known about them, even in theory, that it cannot be ruled out that they would have high spin, an ergoregion and very large mass. 

In short the best present evidence is that a floating orbit which
emits detectable gravitational waves is highly unlikely without the
discovery of new hyper-compact astrophysical objects.

\section{Floating orbits: the ``inverse problem''}

Another way of looking at the question of whether a material body might
permit floating orbits in its spacetime is to consider that we are simply
relaxing some of the assumptions inherent in the Kerr spacetime.
Instead we are considering motion in a spacetime which is axisymmetric and 
stationary but otherwise has arbitrary properties.  
We may ask precisely what qualities such a spacetime structure must have
in order to accomodate floating orbits. This can give us insight into
precisely what properties the central body generating the spacetime must
have to permit floating orbits.

This ``inverse floating orbit problem'' is clearly non-trivial and interesting, even though it cannot be presently 
addressed in full generality. There is a good reason for this: the non-Kerr spacetime we have in mind will be sourced
by a central body whose tidal properties may be difficult to predict. 
The rotational energy of the central body could in principle be extracted (and turned to
orbital energy for the small mass) by some sort of superradiance and/or tidal friction process. Unfortunately our 
understanding of such processes is limited to black holes and does not extend to rapidly rotating relativistic material bodies.  

Despite this setback there is still one avenue in which we can explore the
kinds of spacetime structure which would faciliate floating orbits.
Our proposal is the following: assuming a general axisymmetric and stationary spacetime we can use
the multipole moment formalism developed by Ryan \cite{Ryan97} to calculate the orbital energy radiated to infinity. This can
be balanced by a ``horizon flux'' identical to that for Kerr black holes. Balancing these two fluxes (the requirement for a floating orbit) allows us to 
express the spacetime's quadrupole moment (the first non-trivial deviation with respect to Kerr) in terms
of the mass and spin of the central body and the orbital velocity. Obviously this exercise is not fully consistent but nonetheless
it is not without merits because, as we shall see, it gives us an idea of
the kind of shape a central object should have to make a floating orbit
more possible.

Assuming a test-body in a circular equatorial orbit with (azimuthal) velocity $v$ the GW energy flux lost to infinity 
is given by the following post-Newtonian formula \cite{Ryan97}
\begin{eqnarray}
L_{\infty} &=& L_{Q}
\left[ 1 - \frac{1247}{336} v^2 + 4\pi v^3 - \frac{44711}{9072} v^4  -\frac{11}{4} \frac{S_1}{M^2} v^3 \right. \nonumber \\
&+& \left. \frac{1}{16} \frac{S_1^2}{M^4} v^4 -2 \frac{M_2}{M^3} v^4
+ {\cal O}(v^5)\right]
\end{eqnarray}
where
\begin{equation}
L_{Q}= \frac{32}{5} \left ( \frac{m}{M} \right )^2 v^{10}
\end{equation}
The spacetime we are considering has mass multipoles $\{ M_0, M_2, ...\}$ 
and current multipoles $\{S_1, S_3, ... \}$.
We can identify $M=M_0$ as the central body's mass and $S_1$ as its spin angular momentum. Then $M_2$ is the
quadrupole moment while the remaining higher-order moments only appear in higher post-Newtonian orders.

For the horizon flux and for the level of precision of this calculation we can use the post-Newtonian result 
\cite{Minoetal97}
\begin{equation}
L_{H} = -\frac{1}{4} L_{Q} \frac{S_1}{M^2} \left ( 1+ 3 \frac{S_1^2}{M^4} \right ) v^5
\end{equation}

For a floating orbit with $v=v_f$ we have 
\begin{equation}
\lambda L_{H} + L_{\infty} = 0 
\end{equation}
where the additional parameter $\lambda$ allows us to scale up or down the efficiency of Kerr superradiance. 
We can then obtain a floating orbit solution for $M_2$,
\begin{eqnarray}
\frac{M_2}{M^3}   & \approx &    \frac{1}{2 v^{4}_f} - \frac{1247}{672} \frac{1}{v^{2}_f} + \left ( 2\pi -\frac{11}{8} \frac{S_1}{M^2} \right ) \frac{1}{v_f} \\ &+& \left (  \frac{1}{32} \frac{S_1^2}{M^4} -\frac{44711}{18144} \right ) 
- \frac{\lambda}{8} \frac{S_1}{M^2} \left ( 1 + 3 \frac{S_1^2}{M^4} \right) v_f \nonumber 
\label{M2float}
\end{eqnarray}
For a reasonable range of values such as $v_f = 0.4- 0.7$ and $S_1 = (0.5 - 0.99) M^2$ we find that
floating orbits are possible provided $M_2 \gtrsim 1$ and $\lambda \sim {\cal O}(1)$. 
A positive $M_2$ represents a central body which is a prolate spheroid, a result which does not sit well with the
common expectation of an oblate shape in rotating bodies. For instance, Kerr holes have a quadrupole moment
\begin{equation}
M_2^{\rm Kerr} = -\frac{S_1^2}{M}
\end{equation}
which is always negative, reflecting the oblate shape of the event horizon. There is one remaining hope for those of us of oblate shape:
floating orbit solutions $M_2 < 0$ are possible if accompanied by
amplified horizon flux, i.e. $\lambda \gtrsim 10$ with the same 
parameters as before. Whether such objects can exist is beyond the scope
of this discussion.

The bottom line of this calculation
is rather simple: floating orbits around non-Kerr massive bodies might still exist 
in two scenarios: (i) the spacetime's quadrupole moment has the ``wrong'' 
sign ($M_2 >0$) representing a {\it prolate} rather than an oblate rotating 
body; such an object seems exotic indeed \footnote{In
compact stars any deviation from sphericity is typically dominated
by the rotational flattening, which is oblate in nature.
There are other forces (for instance, magnetic stresses) which 
can deform the body in a prolate manner, but usually the 
deformation is much smaller than the rotational one. 
Systems where the internal stresses (for instance a strong toroidal 
magnetic field) do produce a
prolate shape are unstable and are quickly
driven to a non-axisymmetric configuration where the deformation axis 
and the rotation axis are orthogonal \cite{cutler02}.}
(ii) the quadrupole can be oblate ($M_2 < 0$) but it needs to be 
combined with a ``supra-Kerr''  energy flux removed via superradiance/tidal 
friction by the orbiting body. The likliest possibility for this is an
object which can sustain a relatively high rate of rotation, in the
form of a Kerr parameter significantly greater than
unity. 

\section{Summary and Conclusion}

We examined the possibility of a floating orbit by comparing the orbital flux gained from the spins of the black holes with the orbital flux lost via gravitational wave damping (valid in the adiabatic approximation, applicable in the EMR case discussed in this paper). The flux gained via superradiant scattering from the larger black hole and the flux lost to infinity was found using a code based on superradiant scattering (\cite{gk02}) in the Teukolsky formalism. The flux gained was found to be less than $10\%$ of the flux lost. The flux gained from the smaller black hole was found using calculations by Poisson et al (\cite{poisson09}). This flux was determined to be several orders of magnitude smaller than both the flux lost to infinity and the flux gained from the larger black hole. Therefore, the sum of the fluxes gained from the spins of both black holes still turned out to be less than $10\%$, comfortably less than the $100\%$ required for a floating orbit. 

Adding eccentricity increases the total energy of the orbit. However, simple calculations show that going from circular ($e = 0$) to highly eccentric ($e = 0.99$) orbits increases the energy by a mere $35\%$. Therefore, including eccentricity will likely not increase the flux gained by an order of magnitude, as required if it is to equal the flux lost. Previous work by Hughes \cite{hughes01} suggests that including an angle of inclination will reduce the orbital flux gained, clearly not helping the case of the floating orbit.

The analysis put forward in this paper strongly suggests that EMR binary black holes cannot form floating orbits. Nonetheless, the fact that the orbital flux gained from the spins of the black holess is close to $10\%$ of the flux lost near the ISCO clearly demands that this gain must be taken into account when generating gravitational wave templates from EMRIs. Although we do not consider
comparable mass ratio binaries here, other reports tend to suggest that they
are even less likely than EMRI binaries to exhibit floating 
orbits \cite{Alvi, Lovelace}.

And finally, approximate calculations indicate that floating orbits around non-Kerr objects may be possible provided the shape of the central spinning body is prolate rather than oblate, a shape not generally assumed by spinning objects, or that the flux extracted by the orbiting body from the spin of the central body is ``supra-Kerr'', i.e greater than that in the Kerr case studied in this paper.

\begin{acknowledgments}
The authors would like to acknowledge the valuable suggestions of 
Eric Poisson and Scott Hughes as well as an anonymous referee, all of
whom read over the manuscript at various stages. We would also like to
thank Stephen O'Sullivan for many helpful discussions. 
KG is supported by the Ram\'{o}n y Cajal Programme of 
the Spanish Ministerio de Ciencia e Innovaci\'{o}n 
and by the German Science Foundation (DFG) via SFB/TR7.
\end{acknowledgments}

\appendix
\section{The Teukolsky Formalism}\label{Tformalism}
The following is a brief outline of the mathematical formalism used to calculate the horizon flux $L_{H}$ and the flux radiated to infinity $L_{\infty}$ by a particle orbiting a large spinning black hole \cite{teukolskyI73, teukolskyII73, teukolskyIII74}. The central equation of the Teukolsky formalism  (the Teukolsky equation) is a partial differential equation in the perturbed Weyl scalar $\psi_{4}$ \cite{pt72} of the Kerr spacetime of the large central spinning black hole, with the stress-energy tensor of the orbiting particle as the source of the perturbation. This stress-energy tensor is given by \cite{gk02, teukolskyI73, teukolskyII73, teukolskyIII74}: 

\begin{equation}
T^{\alpha\beta} = m\frac{u^{\alpha}u^{\beta}}{\Sigma \sin\theta u^t}\delta(r-r(t))\delta(\theta-\theta(t))\delta(\phi-\phi(t))
\end{equation}
where $u^{\alpha} = (dx^{\alpha})/{d\tau}$ is the four-velocity of the particle, $\Sigma = r^2 +(A\cos\theta)^2$, $A = aM$, and $a$ is the dimensionless spin parameter of the larger black hole.
 
This Teukolsky equation is variable separable in the Boyer-Lindquist coordinates, by assuming $\psi = e^{-i\omega t}e^{im\phi}S(\theta)R(r)$, where $\psi = \rho^{-4}\psi_{4}$ and $\rho = 1/(r-iA\cos(\theta))$. The radial equation in $R(r)$ is \cite{gk02, teukolskyI73, teukolskyII73, teukolskyIII74}: 
\begin{equation}\label{nhradialeq}
\left[\Delta^{2}\frac{d}{dr}\left(\frac{1}{\Delta}\frac{d}{dr}\right)-V(r)\right]R_{lm\omega} = T_{lm\omega}
\end{equation}

with $V(r) = -\frac{K^2+4i(r-M)K}{\Delta}+8i\omega r + \lambda$, $K = (r^2 + A^2) - mA$ and $\Delta = r^2 - Mr + A^2$. $\lambda$ is the eigenvalue 
corresponding to the spin weighted spheroidal harmonics ${_{-2}{S_{lm}(\theta)}}$. 

The asymptotic solutions to the homogeneous part of equation \eqref{nhradialeq} are \cite{gk02, teukolskyI73, teukolskyII73, teukolskyIII74}: 
\begin{equation}\label{routinfty}
 R^{\mathrm{out}}
\left\{
	\begin{array}{ll}
                       R^{\mathrm{out}}(r \rightarrow R_{+}) = C^{in}\Delta^{2}e^{-ikr_{*}} + C^{\mathrm{out}}e^{ikr_{*}} \\
	           R^{\mathrm{out}}(r \rightarrow \infty) =   r^{3}{e^{i\omega r_{*}}} 

	\end{array}
\right.
\end{equation}

\begin{equation}\label{rin+}
 R^{\mathrm{in}}
\left\{
	\begin{array}{ll}
               R^{\mathrm{in}}(r \rightarrow \infty) = B^{\mathrm{in}}e^{-i\omega r_{*}}/r + B^{\mathrm{out}}r^3{e^{i\omega r_{*}}} \\ 
               R^{\mathrm{in}}(r \rightarrow R_{+}) = \Delta^{2}e^{-ikr_{*}} 
	\end{array}
\right. 
\end{equation}   
where $k = \omega - ma/(2R_{+})$.

Using these, the asymptotic solutions to the non-homogeneous equation \eqref{nhradialeq}, with the source term, may be constructed \cite{gk02}:  

\begin{eqnarray}\nonumber
R_{lm\omega}(r \rightarrow R_{+}) &=& \frac{\Delta^2 e^{-ikr_{*}}}{2i\omega B^{\mathrm{in}}}\int_{R_{+}}^{\infty}\frac{T_{lm\omega}(r')R^{\mathrm{out}}(r')}{\Delta^2(r')}dr'\\ &\equiv & Z_{lm\omega}^{\infty}\Delta^{2}(r)e^{-ikr_{*}}
\end{eqnarray}

\begin{eqnarray}\nonumber
R_{lm\omega}(r \rightarrow \infty) &=& \frac{r^3 e^{i\omega r_{*}}}{2i\omega B^{\mathrm{in}}}\int_{R_{+}}^{\infty}\frac{T_{lm\omega}(r')R^{\mathrm{in}}(r')}{\Delta^2(r')}dr'\\ &\equiv & Z_{lm\omega}^{H}r^3 e^{i\omega r_{*}}
\end{eqnarray}

The fluxes $L_{H}$ and $L_{\infty}$ may be calculated using the $Z$ amplitudes defined above. We further define the radial and angular frequencies as $\Omega_{r} = 2\pi/T_r$ and $\Omega_{\phi} = \Delta\phi/T_r$, where $T_r$ is the period of the radial position function $r(t)$ and $\Delta\phi = \phi(t + T_r) - \phi(t)$. The gravitational wave frequency $\omega_{mk}$ is related to a combination of the harmonics of the two orbital frequencies: $\omega_{mk} = m\Omega_{\phi}+ k\Omega_{r} $. We rewrite the $Z$ amplitudes as \cite{gk02}:

\begin{equation}
Z_{lm\omega}^{\infty, H} = \sum_{k = -\infty}^{+\infty}Z_{lmk}^{\infty, H}\delta(\omega-\omega_{mk}) 
\end{equation}

Using these amplitudes, the fluxes are found to be:

\begin{equation}
L_{\infty} = \sum_{l,m,k}^{}\frac{\left |Z_{lmk}\right|^2 }{4\pi\omega_{mk}^2}
\end{equation}

and

\begin{equation}
L_{H} = \sum_{l,m,k}{}\alpha_{lmk}\frac{\left|Z_{lmk}\right|^2}{4\pi\omega_{mk}^2}
\end{equation}

where:

\begin{equation}
\alpha_{lmk} = \frac{256(2mr_{+})^5 k(k^2 + 4\epsilon^2)(k^2 + 16\epsilon^2)\omega_{mk}^3}{C_{lmk}}
\end{equation}

and

\begin{eqnarray}
C_{lmk} = & [& (\lambda + 2)^2 + 4Am\omega_{mk}-4A\omega_{mk}^2] \nonumber \\
                  & \times & (\lambda^2 + 36Am\omega_{mk} - 36A^2\omega_{mk}^2) \nonumber \\ 
                  &+& (2\lambda +3)(96A^2\omega_{mk}^2)\nonumber \\ 
                   &-& 48Am\omega_{mk})+ 144\omega_{mk}^2(M^2 - A^2)\nonumber
\end{eqnarray}

with $\epsilon = \sqrt{M^2-A^2}/4Mr_{+}$. $C_{lmk}$ is the so-called Starobinsky constant.

 
\section{ Absorption of energy and angular momentum by a small BH in the field of a large BH.}
\label{Tfriction}

The content of this appendix is a summary of the work by Poisson and collaborators \cite{poisson04, poisson09} 
on tidally interacting binary black hole systems. The results described below are used in section~\ref{tidal} of the present paper.

We consider a binary black hole system whose members have, respectively, masses $M$ and $m$ and Kerr spin parameters $a$ and $b$.
We shall only be concerned with the extreme mass ratio case $m \ll M$. On the other hand the two spins are arbitrary 
$0 < a < 1$ and $0 < b < 1$ apart from the assumption that they are parallel at all times (aligned or anti-aligned).

Given this set up the small black hole can be placed in a circular equatorial orbit (of radius $R$) in the Kerr spacetime 
of the larger black hole. The orbital velocity and angular frequency are given by the standard expressions
\begin{equation}
v = \left ( \frac{M}{R} \right )^{1/2}, \quad \Omega_{orb} = \pm \frac{v^3}{M(1+av^3)}
\end{equation} 
where the upper (lower) sign corresponds to a prograde (retrograde) orbit with respect to the large
black hole's spin. Two other important kinematical parameters of the system are the horizon frequency $\Omega_{H}$ 
of the small black hole and the angular frequency $\Omega^\dagger$ of the large black hole's tidal field as seen in the rest frame of the small hole. These are given by
\begin{equation}
\Omega_{H} = \frac{b}{2m} ( 1 - \sqrt{1-b^2}  )^{-1}, \qquad   \Omega^\dagger = \pm \frac{v^3}{M}
\end{equation}
Most notably, $\Omega^\dagger$ does not coincide with $\Omega_{orb}$ unless the large hole is non-spinning. 

The principal result of \cite{poisson04, poisson09} is formulae for the flux of energy and angular momentum
absorbed by the small black hole. These can be written in a compact form, reminiscent of the tidal friction formulae of
the Newtonian problem, 
\begin{align}
L_{t} = \dot{m} = {\cal K} \Omega^\dagger (\Omega^\dagger - \Omega_{H} )
\\
\dot{J} = \frac{L_{t}}{\Omega^\dagger} = {\cal K}  (\Omega^\dagger - \Omega_{H} )
\end{align}
where $J = b m^2$ is the small hole's angular momentum and ${\cal K}$ is a quantity constructed by
the large hole's tidal field. The detailed form of this parameter is given by eqn. (9.46) in \cite{poisson04} but is not required here.

The above fluxes are given explicitly by \cite{poisson04, poisson09} in the limit of
``slow rotation'', that is $\Omega_{H} \ll \Omega^\dagger$,
\begin{align}
L_{t} = \frac{32}{5} \left ( \frac{m}{M} \right )^6 v^{18} \Gamma_{S}
\\
\nonumber \\
\dot{J} = \pm \frac{32}{5} \frac{m^6}{M^5}  v^{15} \Gamma_{S}
\end{align}
and in the opposite ``fast rotation'' limit $\Omega_{H} \gg \Omega^\dagger$
\begin{align}
L_{t} = \mp \epsilon \frac{8}{5} \left ( \frac{m}{M} \right )^5 b (1+ 3b^2) v^{15} \Gamma_{K} 
\\
\nonumber \\
\dot{J} = - \epsilon \frac{8}{5} \frac{m^5}{M^4} b (1+ 3b^2)  v^{12} \Gamma_{K}
\end{align}2
where $\epsilon=+1$ ($\epsilon=-1$) for aligned (anti-aligned) spins.
These formulae feature the ``relativistic factor'' $\Gamma$ which, in a loose sense, represents
the distortion of the small hole's effective absorption cross section due to its motion.
As discussed in \cite{poisson09} this factor is always of order unity.

The above fluxes can be either positive or negative; in particular a negative energy flux $L_{t} < 0$ represents
superradiant scattering taking place at the horizon of the small black hole. 
An inspection of the flux formulae reveals that the maximum energy and angular momentum 
that can be removed from the small black hole is the one of the \emph{fast-rotation} limit with \emph{aligned} spins. 
In that case the $\Gamma$-factor is given by
\begin{eqnarray}\label{gammaVba}
\Gamma_{K}(a,b,v) &=& \frac{1-2v^{2}+a^{2}v^{4}}{(1-3v^{2}+2av^{3})^{2}}
\left(1-\frac{4+27b^{2}}{4+12b^{2}}v^{2} \right. \nonumber \\
&-& \left. \frac{4-3b^{2}}{2+6b^{2}}av^{3}+\frac{8+9b^{2}}{4+12b^{2}}a^{2}v^{4}\right)
\end{eqnarray}
Moreover, the same fast-rotation fluxes can be further maximised if the orbit of the small hole is the ISCO
of the large hole's spacetime. This is indeed the scenario considered in section~\ref{tidal}.

\providecommand{\noopsort}[1]{}\providecommand{\singleletter}[1]{#1}%

\end{document}